*Superluminous supernovae: No threat from Eta Carinae*

Brian C. Thomas[1], Adrian L. Melott[2], Brian D. Fields[3], and Barbara J. Anthony-Twarog[2]

1. Department of Physics and Astronomy, Washburn University, 1700 SW College Ave., Topeka, Kansas, 66621, USA; 785-670-2144; brian.thomas@washburn.edu (to whom correspondence should be addressed)
2. Department of Physics and Astronomy, University of Kansas, Lawrence, Kansas, USA
3. Department of Astronomy, University of Illinois, Urbana, Illinois, USA

Recently Supernova 2006gy was noted as the most luminous ever recorded, with a total radiated energy of ~$10^{44}$ Joules. It was proposed that the progenitor may have been a massive evolved star similar to η Carinae, which resides in our own galaxy at a distance of about 2.3 kpc. η Carinae appears ready to detonate. Although it is too distant to pose a serious threat as a normal supernova, and given its rotation axis is unlikely to produce a Gamma-Ray Burst oriented toward the Earth, η Carinae is about 30,000 times nearer than 2006gy, and we re-evaluate it as a potential superluminous supernova. We find that given the large ratio of emission in the optical to the X-ray, atmospheric effects are negligible. Ionization of the atmosphere and concomitant ozone depletion are unlikely to be important. Any cosmic ray effects should be spread out over ~$10^4$ y, and similarly unlikely to produce any serious perturbation to the biosphere. We also discuss a new possible effect of supernovae, endocrine disruption induced by blue light near the peak of the optical spectrum. This is a possibility for nearby supernovae at distances too large to be considered "dangerous" for other reasons. However, due to reddening and extinction by the interstellar medium, η Carinae is unlikely to trigger such effects to any significant degree.

**Keywords:** supernovae, atmosphere, ozone, radiation

1. Introduction

Supernova 2006gy was recently recorded, and is probably the most luminous such event ever observed (Ofek *et al.,* 2007; Smith *et al.,* 2007), about 100 times the optical luminosity of a typical Type II supernova over a longer duration. Smith *et al.* (2007) suggested that the object η Carinae, thought to be an evolved star of order 100 solar masses, may be very similar to the precursor of SN 2006gy. Recent work has suggested that most supernovae are not serious threats to the biosphere due to X-rays or cosmic rays unless they lie within about 10 pc distance (Gehrels *et al.,* 2003), and it has been suggested that a supernova went off at about 30-40 pc some 2.8 My ago (Knie *et al.,* 2004; Fields, 2004; however see Basu *et al.,* 2007) with no major mass extinction events, though possibly some climate perturbation. However, 2006gy is clearly an outlier, and with two orders of magnitude greater luminous energy than a normal supernova, some reconsideration is in order.

In this study we consider the possible effects of a SN2006gy-like event at the distance of η Carinae – 2.3 kpc (Smith, 2006). We look specifically at atmospheric effects of X-rays from the event, including ionization and subsequent changes in atmospheric constituents such as nitrogen oxides and ozone. We also consider possible effects of optical photons on terrestrial organisms. The potential effects of η Carinae producing a gamma-ray burst (GRB) were considered by Dar and De Rujula (2002). However, as these authors also note, even if η Carinae does produce a GRB, the gamma emission is unlikely to be directed at the Earth. The rotation axis of η Carinae is inclined at about 41-42° to the line of sight (Smith, 2006). Given the strong probability that GRB emission would be beamed along the angular momentum axis, it is unlikely that any such emission would include the Earth. Cosmic rays from a GRB or supernova at η Carinae are also a possibility and are discussed below.

2. **Methods**

Our strategy will be to model the effects of 2006gy as if it were at the distance of η Carinae, given their apparent similarity. As noted above, η Carinae is unlikely to affect the Earth as a GRB. We therefore treat it as a probable superluminous supernova.

Supernovae emit electromagnetic energy ($\gamma$), neutrinos ($\nu$), propel a nonrelativistic blast shell of debris, and probably accelerate cosmic rays, primarily consisting of highly relativistic protons (Liegenfelter and Higdon, 2007; and references therein). Although most of the energy is emitted as $\nu$'s, their effect is negligible for any conceivable proximity (Karam, 2002a,b). Furthermore, the blast shell never propagates as far as the 2.3 kpc we are contemplating. Supernovae are a likely source of cosmic rays, but their propagation in the galaxy is essentially diffusive due to interaction of these charged particles with magnetic fields (Büsching *et al.*, 2005). Our galaxy probably averages about 3 supernovae per century; 2006gy is about 100 times the typical luminosity, but the diffusion would smear out cosmic ray arrival times over about $10^4$ years (Dermer and Holmes, 2005). Given the distance and diffusion time, even if the entire energy of the supernova were emitted as cosmic rays, the received energy flux would only be of order $10^{-9}$ J m$^{-2}$ s$^{-1}$. Consequently, at that distance we would not expect a large perturbation in the cosmic ray background due to the event. It has been noted (e.g. Carslaw *et al.*, 2002) that changes in cosmic ray flux may affect climate through low-level cloud formation. While climate change could pose a threat, the connection is controversial (e.g. Sloan and Wolfendale 2007), and the change in flux is not likely to be large. We will model the effect of the X-ray photons alone.

We performed our modeling using the Goddard Space Flight Center two-dimensional (latitude, altitude) time-dependent atmospheric model that has been used extensively to study the effects of solar flares, as well as supernovae and gamma-ray burst effects. We will describe the code only briefly, given accounts elsewhere (see Thomas *et al.,* 2005; Ejzak *et al.,* 2007; Thomas *et al.,* 2007; and references therein). There are 18 bands of latitude and 58 log pressure bands. The model computes atmospheric constituents in a largely empirical background of solar radiation variations, with photodissociation, and including winds and small scale mixing.

We have placed our source at latitude -60°, corresponding to the location of η Carinae 60° south of the celestial equator. We have timed the outbreak to occur at either of the equinoxes and assume 200 day duration. Given the nature of our conclusion, the time dependence is not important. Due to differences in photodissociation effects, these should be quite close to representing the minimum and maximum atmospheric chemistry perturbations. Reddening toward η Carinae is E(B-V) ~ 0.4 to 0.5 (Davidson and Humphreys, 1997) which would correspond to an X-ray absorbing column of $n_H$ ~ 2-2.6 × $10^{-21}$ cm$^{-2}$ (Prendehl and Schmidt, 1995); this has X-ray absorbing effects (Yao *et al.,* 2006) that are negligible for our purposes.

Atmospheric ionization is computed in a separate off-line code and then used in the atmospheric model as a source of nitrogen oxides (denoted $NO_y$; most importantly NO and $NO_2$). For these computations we adopt a Raymond-Smith thermal plasma spectrum with kT = 1 keV in order to approximate the X-ray photon flux spectrum (following the procedure described in Smith *et al.*, 2007). The luminosity spectrum has the form

$$\frac{dL_x}{dE} = L_0 e^{-E/kT}$$

where *E* is the photon energy. The total luminosity in the X-ray between 0.5 and 2 keV as given in Smith *et al.* (2007) is $L_x = 1.65 \times 10^{32}$ J s$^{-1}$. Integrating over this band we determine the value of $L_0$ to be $3.50 \times 10^{32}$ J s$^{-1}$ keV$^{-1}$.

We have not included the unobserved shock breakout X-rays which we estimate to be about $10^{40}$ J at an equivalent temperature of 0.3 keV. Although this is greater than the integrated X-ray energy from the extended emission, it is even softer photons which we know have less effect on the atmosphere (Ejzak *et al.*, 2007). In addition, they may have been almost totally absorbed and/or downscattered by the robust circumstellar medium of such an object (Davidson and Humphreys 1997), the mass of which has been estimated as 12.5 solar masses (Smith *et al.*, 2003).

The differential photon flux spectrum is then given by

$$\frac{dN}{dE} = A E^{\alpha} e^{-E/kT}$$

where $\alpha = -1.0$ and $A = L_0/(4\pi D^2)$ is the flux constant for the event at distance D in centimeters. ($L_0$ is converted from J s$^{-1}$ keV$^{-1}$ to s$^{-1}$ to yield units for dN/dE of number of photons cm$^{-2}$ s$^{-1}$ keV$^{-1}$.)

Following the procedure used by our group in modeling GRBs, the photon flux in each energy bin in the range 0.1 to 10 keV is obtained by integrating the above spectrum for that bin. The photon flux is propagated through a standard atmosphere (adjusted for appropriate latitude and time when input to the atmospheric model) and is attenuated with altitude by an exponential decay law with energy-dependent coefficients taken from a lookup table. This data is taken from Plechaty *et al.* (1981) in the range 0.1 to 1 keV and from the NIST XCOM database (Berger *et al.*, 2005) in the remainder of the range to 10 keV (see Ejzak *et al.*, 2007 for a more detailed discussion). Finally, the atmospheric ionization rate is determined from the photon energy flux in a given altitude layer divided by 35 eV, the average energy required to produce one ion pair in air (Porter *et al.*, 1976). A more detailed discussion of this procedure is given in Thomas *et al.* (2005).

Photons with sufficient energy ionize and dissociate molecules in the atmosphere. The dominant effect is the resultant synthesis of various oxides of nitrogen. Two of these, NO and $NO_2$, persist in the atmosphere for years, and catalyze the destruction of ozone ($O_3$). Stratospheric ozone absorbs strongly in the UVB (280-315 nm) thus protecting the biosphere from damaging radiation. UVB is implicated in a wide variety of severe biological damage, including to phytoplankton, which lie at the base of the food chain in the oceans. Other effects, such as nitrate deposition at the surface and reduced solar irradiation due to absorption by $NO_2$ are less significant (Melott *et al.*, 2005), so we will first examine ozone depletion as a consequence of X-ray emission from our model superluminous supernova.

It is worth noting that η Carinae currently emits hard X-rays (Corcoran *et al.*, 1998; Hamaguchi *et al.*, 2007). However, the received fluence is about $5 \times 10^{-7}$ J m$^{-2}$, which is approximately 11 orders of magnitude too small to produce significant atmospheric effects (Thomas *et al.*, 2005, Ejzak *et al.*, 2007).

3. **Simulation results**

We find minimal atmospheric effects due to the supernova X-rays received. Atmospheric ionization is limited to altitudes above about 70 km and the maximum volume ionization rate is about 0.01 ions cm$^{-3}$ s$^{-1}$. We have modeled effects for an event at the equinoxes in late March and late September. We have examined changes in nitrogen oxides (NO$_y$) and ozone (O$_3$) and find somewhat greater changes for the March case. Maximum change in these constituents occurs about three months after the start of the event for the March case, and about eight months after for the September case. In both cases the changes in atmospheric constituents are limited to altitudes above about 50 km and latitudes southward of about -40°. Changes in constituents extend to altitudes below that of the direct ionization due to downward transport. However, the changes are confined to altitudes well above the stratosphere and so have almost no impact on total column ozone.

Changes in NO$_y$ and O$_3$ can be examined in several ways. We compare values for a model run that includes ionization from the supernova X-rays to a run without this perturbation. From the point of view of an observer on the ground, column density changes are the most important, especially when considering changes in $O_3$. We find globally averaged decrease in column density of $O_3$ of about 1 part per million at most (March case). The fractional change in column density $O_3$ at specific latitudes is about $10^{-4}$ at maximum (over latitude -85°, for the March case).

We have also examined changes in constituent number density at specific locations in altitude and latitude, at times when the effects are at a maximum. Three months after the start of the event in late March, $NO_y$ has *increased* by about 67% maximum, at an altitude of about 85 km and latitude of -85° (see Figure 1). The maximum $O_3$ *decrease* for this case is in the same region and is about 1.4% (see Figure 2). Similar but somewhat smaller changes are found about eight months after the start of the event in late September, with an increase in $NO_y$ of about 10% and a decrease in $O_3$ of about 0.4%.

Differences in timing and magnitude of these changes for the events in March and September are due to differences in the presence and intensity of sunlight at high southern latitudes at different times of year. The ionization is almost completely contained within the southern hemisphere and atmospheric transport concentrates the nitrogen oxides produced to high southern latitudes fairly quickly. The absence of sunlight enhances the efficiency of ozone depletion by $NO_y$ compounds due to the lack of competing photolytic reactions. In particular, for the March case the event occurs at the start of south polar fall and so sunlight is becoming less intense in subsequent months, allowing greater and more rapid depletion. Conversely, for the September case the event occurs at the start of south polar spring and maximum depletion is "delayed" until about eight months later. In addition, since the event lasts just over six months and south polar night is setting in just after the event has ended, the atmosphere has started to recover (by removing $NO_y$ compounds) and therefore the overall ozone depletion is less severe.

It is important to note that SN 2006gy was unusually X-ray faint, as much as 1000 times fainter than might be expected. In order to more securely assess the potential threat from η Carinae producing an analogous event, we have also modeled the atmospheric effects for an event with the same characteristics but 1000 times greater X-ray luminosity. We find that the atmospheric effects are larger, as would be expected. However, the maximum reduction in column density $O_3$ at a given latitude and time is only about 1.5%. This is smaller than the current globally averaged depletion in $O_3$ column density of a few percent (WMO, 2003). This case should represent an upper limit on the likely X-ray luminosity – we therefore conclude that even this extreme case would not constitute a significant threat.

The fact that we did not include time dependence of the X-ray emission and did not include any possible effect of X-rays at shock breakout probably together introduces a factor of 2 uncertainty into the total atmospheric ionization, since it scales less than linearly with fluence (Thomas *et al.,* 2005; Ejzak *et al.,* 2007). Given the nature of our results, this is unimportant.

The unusually low X-ray luminosity of SN 2006gy combined with its unusually high optical luminosity motivates more attention to possible effects of optical photons, which follows.

4. **Effects of optical photons**

One cannot dismiss the effect of optical photons, even though they do not ionize the atmosphere. The optical spectrum of SN 2006gy peaked at about 400 nm, in the blue. This is important in considering the effects of optical photons, because there are strong indications that even short-time exposure to low levels of short-wavelength optical light (~450 nm) can strongly affect the human endocrine system (Cajochen *et al.*, 2004), as well as that of other mammals (e.g. Sinhasane and Joshi, 1998; Aral *et al.*, 2006). This is a known risk factor for mood disorders (Nurnberger *et al.*, 1988) and more significantly, probably via melatonin suppression, cancer (Reiter *et al.*, 2006). Melatonin suppression can also contribute to insomnia, a risk factor for infection, although effects on the immune system are not confined to mammals (Shirasu-Hiza *et al.*, 2007). For a review, see Brainard and Hanifin (2005).

With a luminosity of about $3 \times 10^{37}$ J s$^{-1}$, about 0.5 mW m$^{-2}$ could irradiate the Earth's surface for several months over much of the southern hemisphere. This is well below the brightness of the full Moon, but is much more strongly peaked in the blue, and can be compared with values used in some biological studies of light pollution (for a capsule review see Reiter *et al.*, 2006). The reason for greater effectiveness of blue light is not known, but evidence points to a non-rod, non-cone photoreceptor (Thapan *et al.*, 2001). However, there is about 1.7 magnitude of visual extinction between here and η Carinae (Davidson and Humphreys 1997), which would reduce the optical light intensity to about 20% of its initial value. Since the scattering of photons by interstellar dust is greater for short wavelengths, the blue component would be reduced even more than this. We estimate reduction to about 12% at 440 nm and 10% at 400 nm. Optical irradiation is therefore probably not important in this case, but we summarize some of the considerations below for future evaluation of effects of other nearby supernovae.

Seasonal and latitudinal effects are important. Presumably for the optical effects, the object would have to be well above the horizon at night in order to affect sleeping animals. η Carinae could be well above the horizon at night for much of the southern hemisphere. For any observer south of about -30°, η Carinae is always above the horizon. The only large land masses that meet this criterion are Antarctica, New Zealand, and extreme southern Australia and South America. However, even at the equator, it would be 30° above the horizon at midnight part of the year: February-March. The photobiological effect would presumably be greatest then. The effect of blue optical photons has not been previously considered as an effect of supernovae or other astrophysical events on terrestrial lifeforms. Both the atmospheric chemistry effects of ionizing photons and the optical effects have a seasonal and latitudinal component which modulates the probable biological effects.

For a nearer supernova, this seasonal interaction between ionization and photobiological effects might be an additional new consideration in evaluating their probable biological impact. Furthermore, with smaller distances, reddening would be negligible. A more quantitative estimate can be used with a threshold around 0.1 W m$^{-2}$ (Brainard *et al.,* 1984) for "cool white" light, and an approximate luminosity of $3 \times 10^{35}$ J s$^{-1}$ for a "typical" supernova, the biological threshold would be reached at about 30 pc, somewhat greater than distances for major conventional considerations of atmospheric ionization; an event at this distance would be expected roughly once per 20 My on average (Gehrels *et al.,* 2003). Clearly the effect of blue optical photons may be competitive, and needs to be added to the evaluation of possible effects of "generic" supernovae on the biosphere.

It is possible that UV emission from such a supernova could be important at the Earth's surface, especially UVA, which is not strongly blocked by ozone. However, for the case we are considering here, any UV emission from η Carinae would be strongly attenuated by the circumstellar medium. Therefore, we do not expect any significant UV irradiation from the supernova event itself and have not considered any such effects.

4.   **Discussion**

It is interesting to estimate how close a SN 2006gy-like event would have to be to induce serious, extinction-level atmospheric effects due to the keV X-rays. Using guidelines developed by Ejzak *et al.* (2007), we estimate that for keV photons, deposition of about 1 MJ m$^{-2}$ would be needed for about one-third depletion of atmospheric ozone, the catastrophic level. Such an event would need to be about 0.3 pc away—extremely unlikely to occur in the lifetime of the Earth. This relatively short required distance is a consequence of the very soft X-ray spectrum of SN 2006gy, which we have assumed for an event at η Carinae as well. A soft X-ray spectrum reduces the intensity of atmospheric effects since the photons have a larger cross section and interact high in the atmosphere, well above the protective ozone layer in the stratosphere. Ozone depleting compounds produced at these high altitudes are largely destroyed by photolysis before they can be transported downward. It is possible that the X-ray spectrum could be harder for an SN 2006gy-like event. A much harder X-ray spectrum would change this distance to perhaps a few parsecs – still rather unlikely. If we consider the optical component, an event at about 3 pc would begin to exceed 10% of solar luminosity. Therefore, for an event within a few parsecs the optical luminosity could, depending on the hardness of the X-ray spectrum, have much more important consequences.

We have two important conclusions: (1) According to our estimates, η Carinae is not likely to have any serious impacts on the terrestrial biosphere, even if it emerges as a superluminous supernova, an analog of SN 2006gy. (2) For ordinary supernovae, the endocrine disruptive effects of blue light may be a dangerous effect for some terrestrial animals over a large part of the globe with an average frequency of 20 million years. This possible effect needs further study.

## 5. Acknowledgments

We acknowledge a Washburn University Small Research Grant, as well as support from NASA Astrobiology: Exobiology and Evolutionary Biology Program under grants number NNG04GM14G and EXB030000031. We thank Arlin Crotts and Marc Sher for valuable personal communications. We also thank Nathan Smith and two anonymous referees for helpful comments.

**References**

Aral, J. *et al.* (2006) Response of the Pineal Gland In Rats Exposed to Three Different Light Spectra of Short Periods, *Turk. J. Anim. Sci.* 30, 29-34.

Basu, S., Stuart, F.M., Schnabel, C., and Klemm, V. (2007) Galactic-Cosmic-Ray-Produced $^3$He in a Ferromanganese Crust: Any Supernova $^{60}$Fe Excess on Earth? *Phys. Rev. Lett.* 98, 141103, doi: 10.1103/PhysRevLett.98.141103

Berger, M.J. *et al.* (2005) XCOM: Photon Cross Section Database, Version 1.3 (Gaithersburg: NIST), http://physics.nist.gov/xcom

Brainard, G.C. *et al.* (1984) The Influence of various irradiances of artificial light, twilight, and moonlight on the suppression of pineal melatonin content in the Syrian Hamster, *J. Pineal Research* 1, 105-119.

Brainard, G.C. and Hanifin, J.P. (2005) Photons, Clocks, and Consciousness, *J. Biological Rhythms* 20, 314-325. DOI: 10.1177/0748730405278951

Büsching, I. *et al.* (2005) Cosmic-Ray Propagation Properties for an Origin in Supernova Remnants, *Astrophys. J.* 619, 314-326. doi: 10.1086/426537

Cajochen, C. *et al.* (2004) High Sensitivity of Human Melatonin, Alertness, Thermoregulation, and Heart Rate to Short Wavelength Light, *J. Clinical Endocrinology & Metabolism* 90, 1311-1316. doi:10.1210/jc.2004-0957

Calzavara, A.J. and Matzner, C.D. (2004) Supernova properties from shock breakout X-rays, *MNRAS* 351, 694-706.

Carslaw, K.S. *et al.* (2002) Cosmic rays, Clouds, and Climate, *Science* 298, 1732-1737.

Corcoran, M.F., Petre, R., Swank, J.H. and Drake, S.A. (1998) The ASCA X-Ray Spectrum of η Carinae, *Astrophys. J.* 494, 381-395.

Dar, A. and De Rujula, A. (2002) The threat to life from Eta Carinae and gamma-ray bursts, in *Astrophysics and Gamma Ray Physics in Space,* ed. A. Morselli and P. Picozza (Frascati Physics Series Vol. XXIV), pp. 513-523.

Davidson, K. and Humphreys, R.M. (1997) Eta Carinae and Its Environment, *Ann. Rev. Astron. Astrophys.* 35, 1-32.

Dermer, C.D. and Homes, J.M. (2005) Cosmic Rays from Gamma-Ray Bursts in the Galaxy, *Astrophys. J.* 628, L21-L24.

Ejzak, L.M., Melott, A.L., Medvedev, M.V., and Thomas, B.C. (2007) Terrestrial Consequences of Spectral and Temporal Variability in Ionizing Photon Events, *Astrophys. J.* 654, 373-384.

Fields, B.D. (2004) Live radioisotopes as signatures of nearby supernovae, *New Astronomy Reviews* 48, 119-123 doi:10.1016/j.newar.2003.11.017

Gehrels, N. *et al.* (2003) Ozone Depletion from Nearby Supernovae, *Astrophys. J.* 585, 1169-1176. doi: 10.1086/346127

Hamaguchi, K. *et al.* (2007) X-Ray Spectral Variation of η Carinae through the 2003 X-Ray Minimum, *Astrophys. J.* 663, 522-542.

Karam, P.A. (2002a) Gamma and neutrino radiation does from gamma ray bursts and nearby supernovae, *Health Physics* 82, 491-499.

Karam, P.A. (2002b) Terrestrial radiation exposure from supernova-produced radiactivities, *Radiation Physics and Chemistry* 64, 77-87.

Knie, K. *et al.* (2004) $^{60}$Fe Anomaly in a Deep-Sea Manganese Crust and Implications for a Nearby Supernova Source, *Phys. Rev. Lett.* 93, 171103 doi:10.1103/PhysRevLett.93.171103

Ligenfelter, R.E. and Higdon, J.C. (2007) Cosmic Rays, Dust, and the Mixing of Supernova Ejecta into the Interstellar Medium in Superbubbles, *Astrophys. J.* 660, 330-335. doi: 10.1086/513420

Nernberger, J.I. *et al.* (1988) Supersensitivity to melatonin suppression by light in young people at high risk for affective disorder. *Neuropsychopharmacology* 1, 217-223.

Ofek, E.O. *et al.* (2007) SN 2006gy: An extremely luminous supernova in the galaxy NGC 1260, *Astrophys. J. Lett.* 659, L13-L16.

Pittard, J.M. (2003) Enigmatic Eta Carinae, *Astronomy & Geophysics* 44, 1.17-1.22, doi:10.1046/j.1468-4004.2003.44117.x

Plechaty, E.F., Cullen, D.E., and Howerton, R.J. (1981) Tables and Graphs of Photon-Interaction Cross Sections from 0.1 keV to 100 MeV Derived from the LLL Evaluated-Nuclear-Data Library (Berkeley: LLNL)

Porter, H.S., Jackman, C.H., and Green, A.E.S. (1976) Efficiencies for production of atomic nitrogen and oxygen by relativistic proton impact in air, *J. Chem. Phys.* 65, 154-167.

Predehl, P. and Schmitt, J. H. M. M. (1995) X-raying the interstellar medium: ROSAT observations of dust scattering halos, *Astron. Astrophys.* 293, 889-905

Reiter, R.J., Gultekin, F., Manchester, L.C., and Tan, D.-X. (2006) Light Pollution, melatonin suppression, and cancer growth, *J. Pineal Res*. 40, 357-358.

Shirasu-Hiza, M.M., Dionne, M.S., Pham, L.N., Ayers, J.S., and Schneider, D.S. (2007) Interactions between circadian rhythm and immunity in Drosophila melanogaster, *Current Biology* 17, R353-R355. doi:10.1016/j.cub.2007.03.049

Sinhasane, S.V. and Joshi, B.N. (1998) Exposure to Different Spectra of Light, Continuous Light and Treatment with Melatonin Affect Reproduction in the Indian Desert Gerbil Meriones hurrianae (Jerdon), *Biological Signals and Receptors* 7, 179-187. DOI: 10.1159/000014543

Sloan, T. and Wolfendale, A.W. (2007) Cosmic Rays and Global Warming, submitted to 30[th] International Cosmic Ray Conference, preprint: http://arxiv.org/abs/0706.4294

Smith N, *et al.* (2003) Mass and Kinetic Energy of the Homunculus Nebula around η Carinae, *Astronomical J.* 125, 1458-1466.

Smith, N. (2006) The Structure of the Homunculus. I. Shape and Latitude Dependence from H2 and [Fe ii] Velocity Maps of η Carinae, *Astrophys. J.* 644, 1151-1163.

Smith, N. *et al.* (2007) SN 2006gy: Discovery of the most luminous supernova ever recorded, powered by the death of an extremely massive star like Eta Carinae, *Astrophys. J.* in press (astro-ph/0612617)

Thapan K., Arendt, J. and Skene, D.J. (2001) An action spectrum for melatonin suppression: evidence for a novel non-rod, non-cone photoreceptor system in humans, *J. Physiology* 535, 261-267.

Thomas, B.C. *et al.* (2005) Gamma-Ray Bursts and the Earth: Exploration of Atmospheric, Biological, Climatic, and Biogeochemical Effects, *Astrophys. J.* 634, 509-533. doi: 10.1086/496914

Melott, A.L., Thomas, B.C., Hogan, D.P., Ejzak, L.M., and Jackman, C.H. (2005) Climatic and Biogeochemical Effects of a Galactic Gamma-Ray Burst, *Geophysical Research Lett.* 32, L14808. doi: 10.1029/2005GL023073

Thomas, B.C., Jackman, C.H., Melott, A.L. (2007) Modeling atmospheric effects of the September 1859 Solar flare, *Geophysical Research Lett.* 34, L06810. doi: 10.1029/2006GL029174

WMO (World Meteorological Organization) (2003), Scientific Assessment of Ozone Depletion: 2002, Global Ozone Research and Monitoring Project – Report No. 47, Geneva.

Yao, W.-M. *et al.* (2006) Review of Particle Physics, *J. Phys. G: Nucl. Part. Phys*. 33, 1-1232.  doi:10.1088/0954-3899/33/1/001

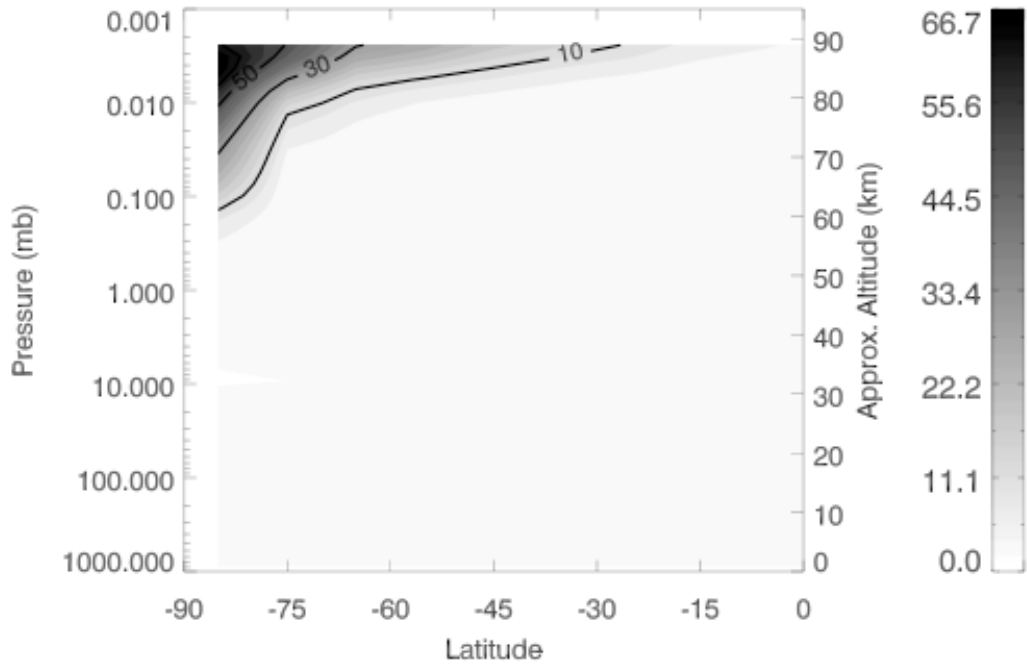

Figure 1

Percent difference in profile $NO_y$ between perturbed and unperturbed run for the March case, at three months after the start of the event. Contour lines are at 10, 30, 50, and 60 percent.

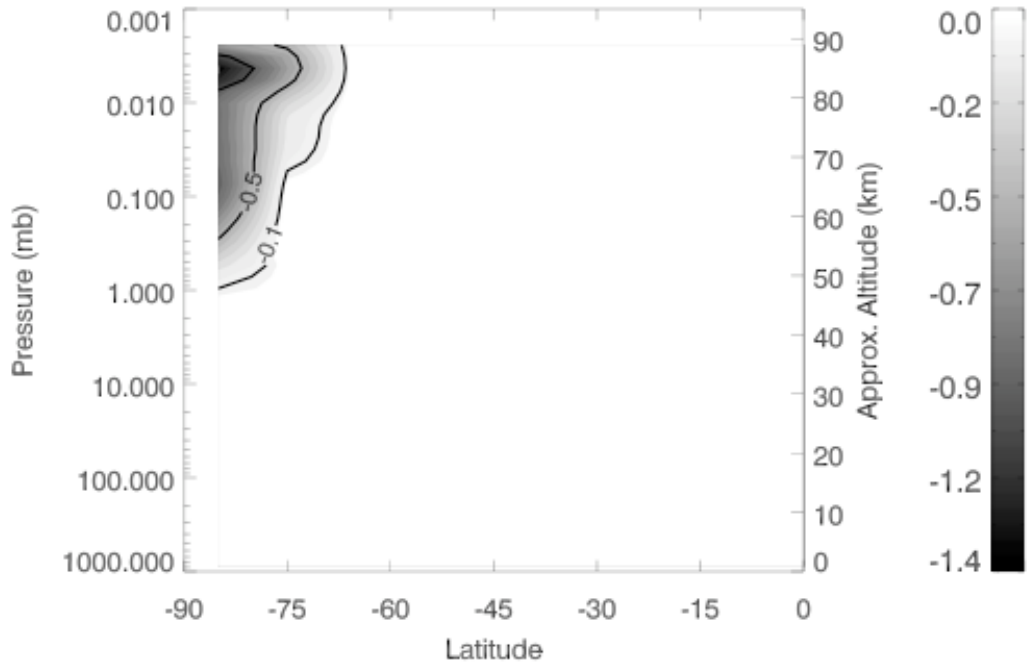

Figure 2

Percent difference in profile $O_3$ between perturbed and unperturbed run for the March case, at three months after the start of the event. Contour lines are at -0.1, -0.5, and -1.0 percent.